\documentstyle[11pt,newpasp,twoside,epsf]{article}
\newcommand{\CDM}{{\rm {\scriptscriptstyle {CDM}}}}
\def\edcomment#1{\iffalse\marginpar{\raggedright\sl#1\/}\else\relax\fi}
\reversemarginpar
\begin{document}
\title{Simulations of Large-Scale Structure \\ in the New Millennium}
 \author{Yasushi Suto}
\affil{Department of Physics, School of Science, The University of
    Tokyo, Tokyo 113-0033, Japan}
\begin{abstract}
Simulations of large-scale structure in the universe have played a vital
role in observational cosmology since 1980's in particular. Their
important role will definitely continue to be true in the 21st
century. Rather the requirements for simulations in the precision
cosmology era will become more progressively demanding; they are
supposed to fill the missing link in an accurate and reliable manner
between the ``initial'' condition at z=1000 revealed by WMAP and the
galaxy/quasar distribution at z=0 - 6 surveyed by 2dF and SDSS. In this
review, I will summarize what we have learned so far from the previous
cosmological simulations, and discuss several remaining problems for the
new millennium.
\end{abstract}

\section{Introduction: evolution of cosmological simulations}

Cosmological $N$-body simulations started in late 1970's, and since then
have played an important part in describing and understanding the
nonlinear gravitational clustering in the universe.  As far as I know,
the cosmological N-body simulation in a comoving periodic cube, which is
quite conventional now, was performed for the first time by Miyoshi \&
Kihara (1975) using $N=400$ particles.  Figure 1 plots the evolution of
the number of particles employed in cosmological N-body
simulations. Here I consider only the ``high-resolution'' simulations
including Particle-Particle, Particle-Particle--Particle-Mesh, and tree
algorithms which are published in refereed journals (excluding, e.g.,
conference proceedings). I found that the evolution is well fitted by
%%%%%%%%%%%%%%%%%%%%%%%%%%%%%%%%%%%%%%%%%%%%%%%%%%%%%%%%%%%%
\begin{eqnarray}
\label{eq:Nyear}
 N= 400 \times 10^{0.215({\rm Year}-1975)} ,
\end{eqnarray}
%%%%%%%%%%%%%%%%%%%%%%%%%%%%%%%%%%%%%%%%%%%%%%%%%%%%%%%%%%%%
where the amplitude is normalized to the work of Miyoshi \& Kihara
(1975).  Just for comparison, the total number of CDM particles of mass
$m_{\rm CDM}$ in a box of the universe of one side $L$ is
%%%%%%%%%%%%%%%%%%%%%%%%%%%%%%%%%%%%%%%%%%%%%%%%%%%%%%%%%%%%
\begin{eqnarray}
 N= \frac{\Omega_\CDM \rho_{\rm cr} L^3}{m_\CDM}
\approx 10^{83} \left(\frac{\Omega_\CDM}{0.23} \right)
\left(\frac{L}{1 h^{-1}{\rm Gpc}}\right)^3
\left(\frac{1 {\rm keV}}{m_\CDM}\right)
\left(\frac{0.71}{h}\right) .
\end{eqnarray}
%%%%%%%%%%%%%%%%%%%%%%%%%%%%%%%%%%%%%%%%%%%%%%%%%%%%%%%%%%%%
If I simply extrapolate equation (1) and adopt the WMAP parameters
(Spergel et al. 2003), then the number of particles that one can
simulate in a $(1 h^{-1}{\rm Gpc})^3$ box will reach the real number of
CDM particles in December 2348 and February 2386 for $m_\CDM= 1{\rm
keV}$ and $10^{-5}{\rm eV}$, respectively. I have not yet checked the
above arithmetic, but the exact number should not change the basic
conclusion; simulations in the new millennium will be {\it unbelievably}
realistic.

%%%%%%%%%%%%%%%%%%%%%%%%%%%%%%%%%%%%%%%%%%%%%%%%%%%%%%%%%%%%%%%%
\begin{figure}[tbh]
\begin{center}
\leavevmode\epsfxsize=7.50cm \epsfbox{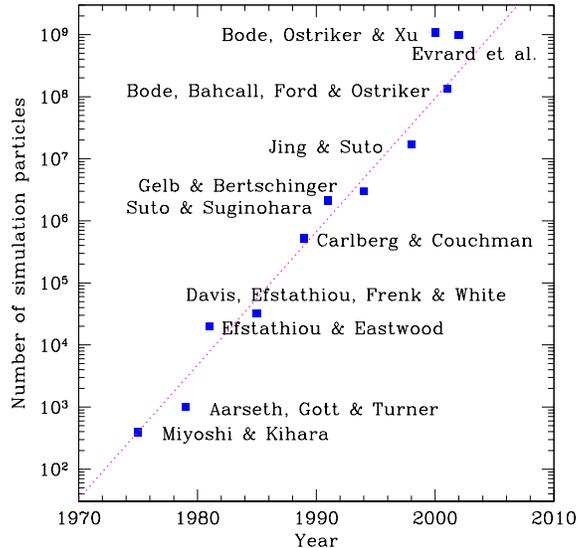} 
\caption{Evolution of the number of particles in ``high-resolution''
cosmological N-body simulations. \label{fig:Nyear}}
\end{center}
\end{figure}
%%%%%%%%%%%%%%%%%%%%%%%%%%%%%%%%%%%%%%%%%%%%%%%%%%%%%%%%%%%%%%%%

\section{1970's: simulating nonlinear gravitational evolution}

It is not easy to identify who attempted seriously for the first time
the numerical simulation of large-scale structure in the universe. Still
I believe that a paper by Miyoshi \& Kihara (1975) is truly pioneering,
and let me briefly mention it here. They carried out a series of
cosmological $N$-body experiments with $N=400$ {\it galaxies}
(=particles) in an expanding universe. The simulation was performed in a
comoving cube with a periodic boundary condition (Fig.2). As the title
of the paper ``{\it Development of the correlation of galaxies in an
expanding universe}'' clearly indicates, they were interested in
understanding why the observed galaxies in the universe exhibit a
characteristic correlation function of $g(r) = (r_0/r)^s$. In fact,
Totsuji \& Kihara (1969) had already found that $r_0=4.7h^{-1}$Mpc and
$s=1.8$ is a reasonable fit to the clustering of galaxies in the Shane
-- Wirtanen catalogue.  One of the main conclusions of Miyoshi \&
Kihara (1975) is that ``{\it The power-type correlation function
$g(r)=(r_0/r)^s$ with $s \approx 2$ is stable in shape; it is generated
from motionless galaxies distributed at random and also from a system
with weak initial correlation }''.

Several years after Totsuji \& Kihara (1969) published the paper,
Peebles (1974) and Groth \& Peebles (1977) reached the same conclusion
independently, which has motivated several cosmological $N$-body
simulations all over the world. Among others, Aarseth, Gott \& Turner
(1979) conducted a series of careful and systematic simulations to
explore nonlinear gravitational clustering.  Those simulations in 1970's
assume that galaxy distribution is well traced by simulations
particles. In fact, the above papers spent many pages in an argument to
justify the assumption, and then attempted to understand the nonlinear
gravitational evolution and to quantitatively describe the large-scale
structure in the computer on the basis of two-point correlation
functions. In this sense, I would say that the simulations in late
1970's are more physics-oriented rather than astronomy.  Also it is
interesting to mention that Prof. Kihara was a solid state physicist in
the University of Tokyo and I speculate that this is why he was able to
accomplish truly pioneering work from such an interdisciplinary point of
view.

%%%%%%%%%%%%%%%%%%%%%%%%%%%%%%%%%%%%%%%%%%%%%%%%%%%%%%%%%%%%%%%%
\begin{figure}[hbt]
\begin{center}
\leavevmode\epsfxsize=6.0cm \epsfbox{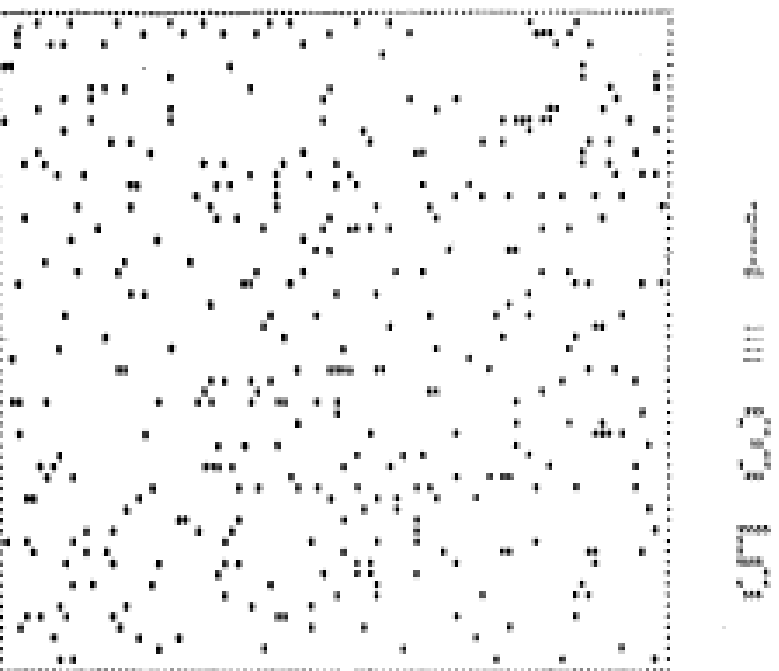} 
\hspace*{0.5cm} 
\leavevmode\epsfxsize=6.0cm \epsfbox{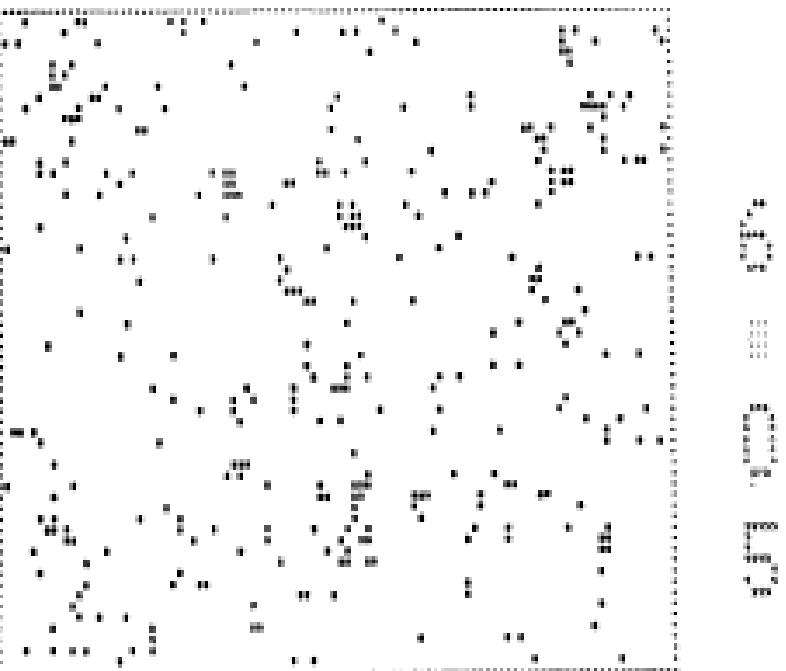}
\vspace*{0.5cm} 
\leavevmode\epsfxsize=6.0cm \epsfbox{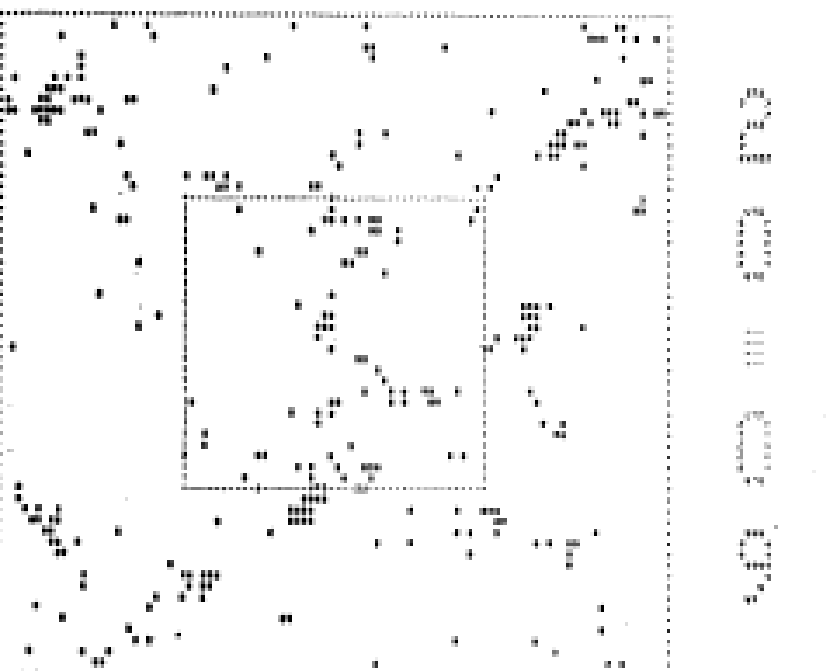} 
\hspace*{0.5cm}
\leavevmode\epsfxsize=6.0cm \epsfbox{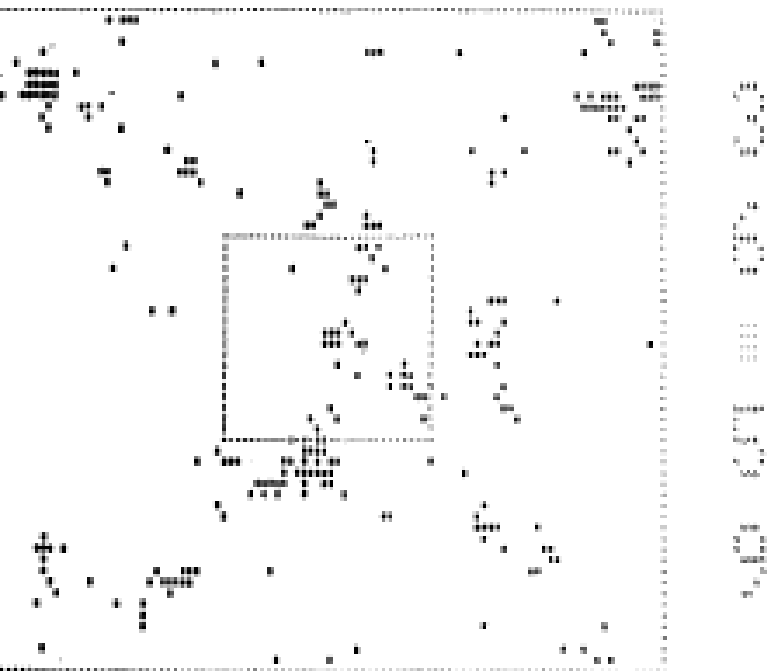} 
\caption{Evolution of clustering from Miyoshi \& Kihara (1975) at
expansion factor (relative to the initial epoch of simulations) of 1.35,
6.05, 20.09 and 36.59.}  \label{fig:MKsimulation}
\end{center}
\end{figure}
%%%%%%%%%%%%%%%%%%%%%%%%%%%%%%%%%%%%%%%%%%%%%%%%%%%%%%%%%%%%%%%%

\section{1980's: introducing galaxy biasing}

The primary goal of simulations of large-scale structure in 1980's was
to predict observable galaxy distribution from dark matter clustering.
Necessarily one had to start distinguishing galaxies and simulation
particles (designed to represent dark matter in the universe), i.e., to
introduce the notion of galaxy {\it biasing} according to the current
terminology. Furthermore a variety of astronomical and/or observational
effects (selection function, redshift-space distortion, etc.) had to be
incorporated towards more realistic comparison with galaxy redshift
survey data which became available those days. Davis, Efstathiou, Frenk
and White (1985) is {\it the} most influential and seminal paper in
cosmological $N$-body simulations in my view.  While their work is quite
pioneering in many aspects, the most important message that they were
able to show in a quite convincing fashion is that simulations of
large-scale structure can provide numerous realistic and testable
predictions of {\it dark} matter scenarios against the observational
data from {\it luminous} galaxy samples.  Considering the fact that they
used only $N=32^3$ particles and thus had to identify the present epoch
as when they advance the simulation merely by a factor of 1.4 relative
to the initial epoch, this presents a convincing case that the most
important is not the quality of simulations but those who interpret the
result.

\section{1990's: more accurate and realistic modeling of galaxy clustering}

I started to work on cosmological N-body simulations around 1987, and it
has been my major research topic for the next several years. At that
time I often asked myself if purely N-body simulations would continue to
advance our understanding of {\it galaxy} clustering significantly. My
personal answer was ``No. Without proper inclusion of hydrodynamics,
radiative processes, star formation and feedback, it is unlikely to
proceed further'', so I moved to more analytical and/or observational
researches. Although I still do not think that my thought was terribly
wrong, I have to admit that my decision was premature; purely N-body
simulations in 1990's turned out to be so successful and they achieved
quite important contributions in (at least) three basic aspects; (i)
accurate modeling of nonlinear two-point correlation functions, (ii)
abundance and biasing of dark matter halos, and (iii) density profiles
of dark halos, which are separately described below.
Thus I returned to simulation work again in late 1990's.

\subsection{Nonlinear two-point correlation function of dark matter}

The first breakthrough came from the discovery of the amazing scaling
property in the two-point correlation functions (Hamilton et al. 1991).
They found that the two-point correlation functions in N-body
simulations can be well approximated by a universal fitting formula
which empirically interpolates the linear regime and the nonlinear
stable solution.  Their remarkable insight was then elaborated and
improved later (e.g., Peacock \& Dodds 1996; Smith et al. 2003),
and the resulting accurate fitting formulae have been applied in a
variety of cosmological analyses.

Figure 3 plots two-point correlations of dark matter from N-body
simulations (Hamana, Colombi \& Suto 2001a). The symbols indicate the
averages over the five realizations from simulations in real space (open
circles) and in redshift space (solid triangles), and the quoted
error-bars represent the standard deviation among them.  The results for
all particles (left panels) agree very well with the theoretical
predictions (solid lines) which combine the Peacock-Dodds formula and
the light-cone effect (e.g., Yamamoto \& Suto 1999; Suto et al. 1999).
The scales where the simulation data in real space become smaller than
the corresponding theoretical predictions simply reflect the force
resolution of the simulations. The result is fairly robust against the
selection effect; Figure 4 indicates that the simulation results and the
predictions are still in good agreement even after incorporating the
realistic selection functions.

One of the main purposes of N-body simulations in 1970's and 1980's was
to compute the nonlinear two-point correlation functions which were
unlikely to predict analytically with a reasonable accuracy. In the
light of this, it is interesting to note that as long as two-point
correlation functions of dark matter are concerned, one does not have to
run N-body simulations owing to the significant progress in
semi-analytical modeling achieved on the basis of previous N-body
simulations.

%%%%%%%%%%%%%%%%%%%%%%%%%%%%%%%%%%%%%%%%%%%%%%%%%%%%%%%%%%%%%%%%%%
\begin{figure}[h]
\begin{minipage}{6.5cm}
\begin{center}
   \leavevmode \epsfxsize=6.5cm \epsfbox{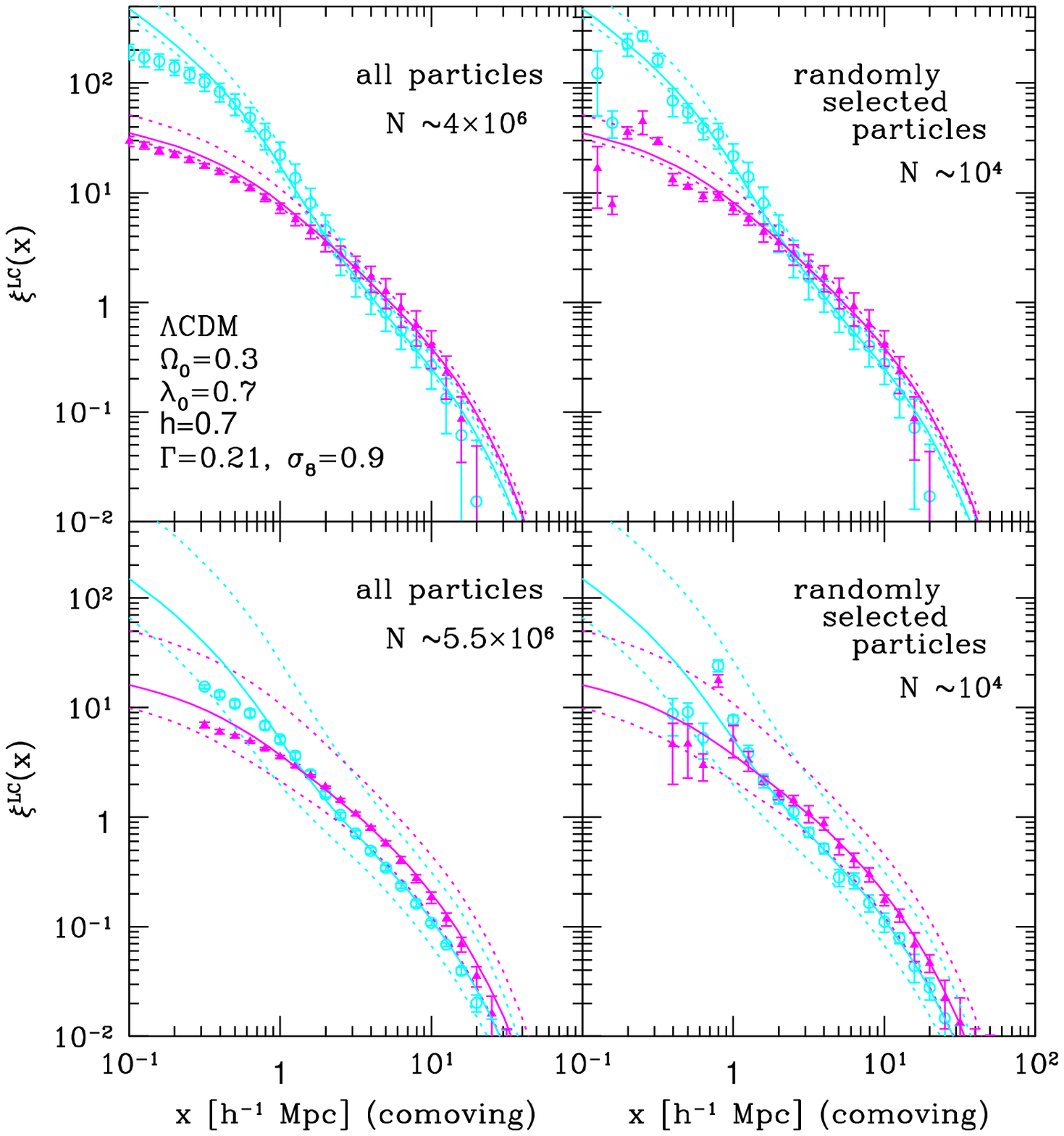} \caption{Two-point
correlation functions of dark matter on the light cone neglecting
selection functions in $\Lambda$CDM model.  {\it Upper:} $z<0.4$, {\it
Lower:} $0<z<2.0$.  {\it Left:} all particles, {\it Right:} randomly
selected particles from the left results (Hamana, Colombi \& Suto
 2001a).}  
\label{fig:xilc_mass_lcdm}
\end{center}
\end{minipage}
\begin{minipage}{6.5cm}
\begin{center}
   \leavevmode \epsfxsize=6.5cm \epsfbox{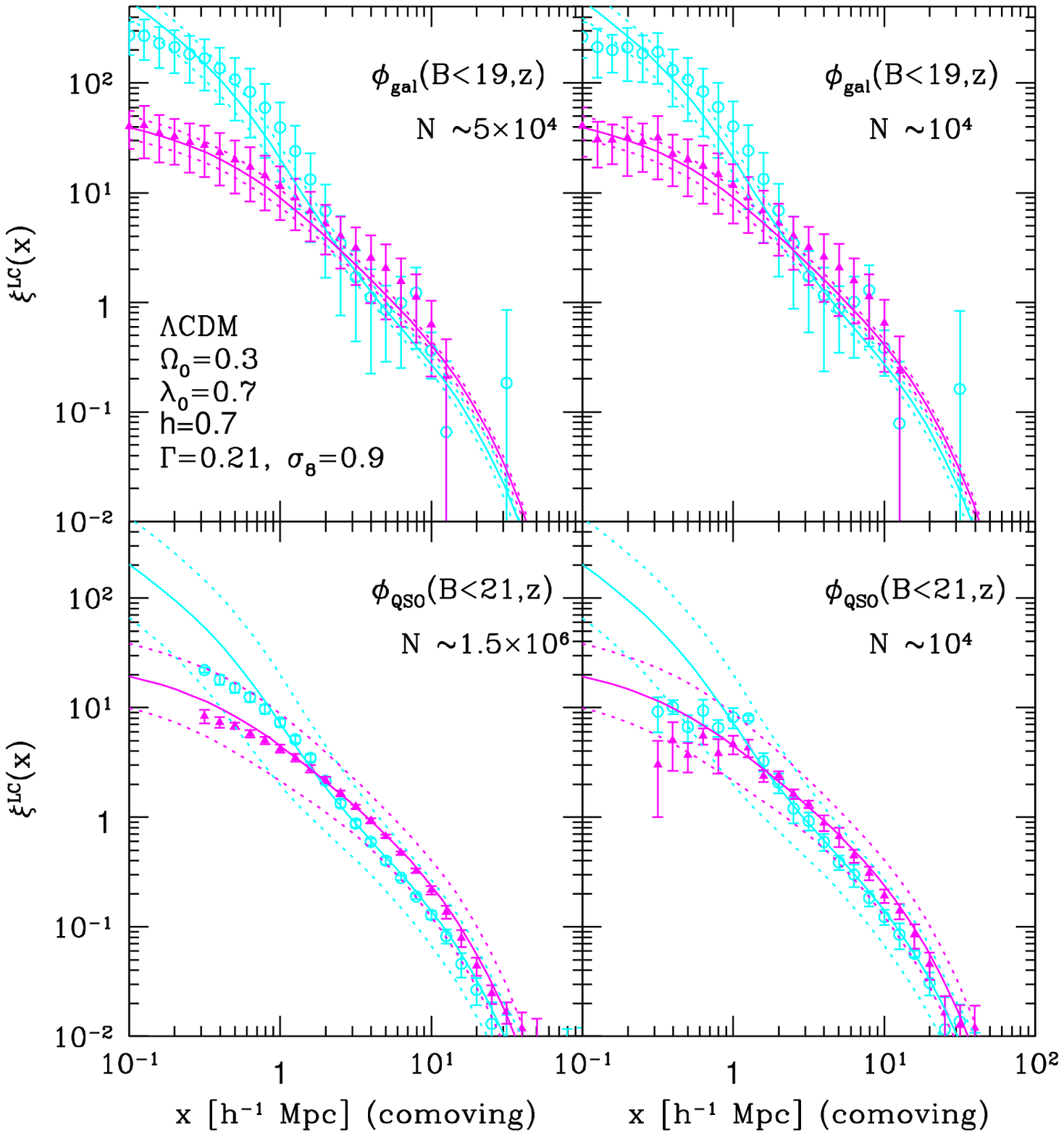} \caption{Same as
Figure 3 but taking account of redshift-dependent selection
functions. The B-band magnitude limits of 19 (upper panels) and 21
(lower panels) are adopted so as to mimic the galaxy and QSO selection
functions (Hamana, Colombi \& Suto 2001a).  }  \label{fig:xilc_phi_lcdm}
\end{center}
\end{minipage}
\end{figure}
%%%%%%%%%%%%%%%%%%%%%%%%%%%%%%%%%%%%%%%%%%%%%%%%%%%%%%%%%%%%%%%%%%%%%

\subsection{Biasing of dark matter halos}

The second remarkable progress where N-body simulations have played a
major role in 1990's is related to the statistics of dark halos, in
particular their mass function and spatial biasing.  The standard
picture of structure formation predicts that the luminous objects form
in a gravitational potential of dark matter halos.  Therefore, a
detailed understanding of halo clustering is the natural next step
beyond the description of clustering of dark matter particles. Both the
extended Press-Schechter theory and high-resolution N-body simulations
have made significant contributions in constructing a semi-analytical
framework for halo clustering. 

For a specific example, let me show our recent mass-, scale-, and
time-dependent halo bias model (Hamana et al. 2001b):
%%%%%%%%%%%%%%%%%%%%%%%%%%%%%%%%%%%%%%%%%%%%%%%%%%%%%%%%%%%%%%%%%%%%%
\begin{eqnarray}
\label{eq:bhalo_MRz}
b_{\rm halo}(M,R,z) &=&
  b_{\scriptscriptstyle\rm ST}(M,z) 
\left[ 1.0+ b_{\scriptscriptstyle\rm ST}(M,z)\sigma_M(R,z) \right]^{0.15}, \\
b_{\scriptscriptstyle\rm ST}(M,z) &=& 1 + \frac{\nu-1}{\delta_c(z)}
+\frac{0.6}{\delta_c(z)(1+0.9\nu^{0.3})},
\end{eqnarray}
%%%%%%%%%%%%%%%%%%%%%%%%%%%%%%%%%%%%%%%%%%%%%%%%%%%%%%%%%%%%%%%%%%%%%
which generalizes the previous work including Mo \& White (1996), Jing
(1998), Sheth \& Tormen (1999) and Taruya \& Suto (2000).  The above
biasing parameter is adopted for $R>2R_{\rm vir}(M,z)$, where $R_{\rm
vir}(M,z)$ is the virial radius of the halo of mass $M$ at $z$, while we
set $b_{\rm halo}(M,R,z)= 0$ for $R<2R_{\rm vir}(M,z)$ in order to
incorporate the halo exclusion effect approximately.  In the above
expressions, $\sigma_M (R,z)$ is the mass variance smoothed over the
top-hat radius $R\equiv (3M/4\pi\rho_0)^{1/3}$, $\rho_0$ is the mean
density, $\delta_c(z) = 3(12\pi)^{2/3}/20D(z)$, $D(z)$ is the linear
growth rate of mass fluctuations, and $\nu=[\delta_c(z)/\sigma_M
(R,z=0)]^2$.
%%%%%%%%%%%%%%%%%%%%%%%%%%%%%%%%%%%%%%%%%%%%%%%%%%%%%%%%%%%%%%%%%%%%%%%%%%
\begin{figure}[htb]
\begin{center}
\leavevmode\epsfxsize=9.0cm \epsfbox{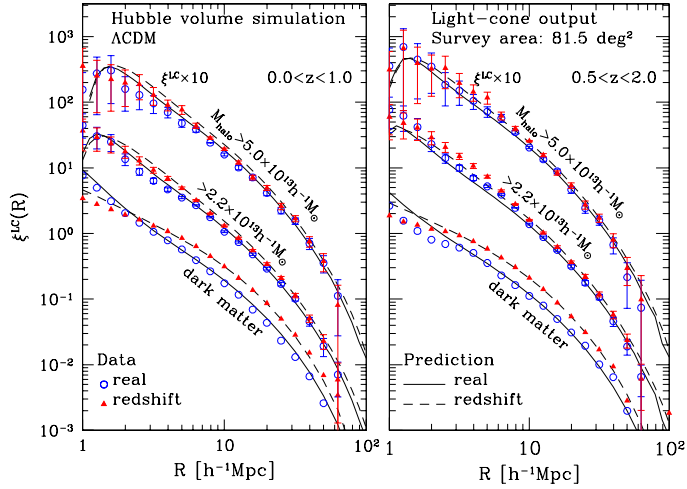}
\caption{Two-point correlation functions of halos on the light-cone;
simulation results (symbols; open circles and filled triangles for real
and redshift spaces, respectively) and our predictions (solid and dotted
lines for real and redshift spaces, respectively).  The error bars
denote the standard deviation computed from 200 random re-samplings of
the bootstrap method.  The amplitudes of $\xi^{LC}$ for $M_{\rm halo}\ge
5.0\times 10^{13}h^{-1} M_\odot$ are increased by an order of magnitude
for clarity (Hamana et al. 2001b).
\label{fig:haloxilc}}
\end{center}
\end{figure}
%%%%%%%%%%%%%%%%%%%%%%%%%%%%%%%%%%%%%%%%%%%%%%%%%%%%%%%%%%%%%%%%%%%%%%%%%%

Figure 5 shows the comparison of the two-point correlation functions of
dark matter halos between the above semi-analytical predictions and the
simulation data (Hamana et al. 2001b).  For this purpose, we analyze
``light-cone output'' of the Hubble Volume $\Lambda$CDM simulation
(Evrard et al. 2002).  The two-point correlation functions on the
light-cone plotted in Figure 5 correspond to halos with
$M>5.0\times10^{13}h^{-1}M_\odot$, $M>2.2\times10^{13}h^{-1}M_\odot$ and
dark matter from top to bottom.  The range of redshift is $0<z<1$ ({\it
Left panel}) and $0.5<z<2$ ({\it Right panel}). Predictions in redshift
and real spaces are plotted in dashed and solid lines, while simulation
data in redshift and real spaces are shown in filled triangles and open
circles, respectively.

Our model and simulation data show quite good agreement for dark halos
at scales larger than $5h^{-1}$Mpc. Below that scale, they start to
deviate slightly in a complicated fashion depending on the mass of halo
and the redshift range.  Nevertheless the clustering of {\it clusters}
on scales below $5h^{-1}$Mpc is difficult to determine observationally
anyway, and our model predictions differ from the simulation data only
by $\sim 20$ percent at most. This illustrates the fact that the
clustering not only of dark matter but also of dark halos, at least as
far as their two-point statistics is concerned, can be described well
semi-analytically without running expensive N-body simulations at all.

\subsection{Density profiles of dark  halos}

The third, and perhaps the most useful in cosmological applications,
result out of N-body simulations in 1990's is the discovery of the
universal density profile of dark halos.

The study of the density profiles of cosmological self-gravitating
systems or dark halos has a long history.  Navarro, Frenk \& White
(1995, 1996, 1997) found that all simulated density profiles can be well
fitted to the following simple model (now generally referred to as the
NFW profile):
%%%%%%%%%%%%%%%%%%%%%%%%%%%%%%%%%%%%%%%%%%%%%%%%%%%%%%%%%%%%%%%%%%%%%
\begin{equation}
\label{eq:nfw}
\rho(r) \propto {1 \over (r/r_{\rm s})(1+r/r_{\rm s})^2}
\end{equation}
%%%%%%%%%%%%%%%%%%%%%%%%%%%%%%%%%%%%%%%%%%%%%%%%%%%%%%%%%%%%%%%%%%%%%
by an appropriate choice of the scaling radius $r_{\rm s}=r_{\rm s}(M)$
as a function of the halo mass $M$.  Subsequent higher-resolution
simulations (Fukushige \& Makino 1997, 2001; Moore et al. 1998; Jing \&
Suto 2000) have indicated that the inner slope of density halos is
steeper than the NFW value, and the current consensus among most N-body
simulators is given by
%%%%%%%%%%%%%%%%%%%%%%%%%%%%%%%%%%%%%%%%%%%%%%%%%%%%%%%%%%%%%%%%%%%%%
\begin{equation}
\label{eq:universal}
\rho(r) \propto {1 \over (r/r_{\rm s})^{\alpha}(1+r/r_{\rm s})^{3-\alpha}}
\end{equation}
%%%%%%%%%%%%%%%%%%%%%%%%%%%%%%%%%%%%%%%%%%%%%%%%%%%%%%%%%%%%%%%%%%%%%
with $\alpha \approx 1.5$ rather than the NFW value, $\alpha=1$ for
$r>0.01r_{\rm vir}$.

Actually it is rather surprising that the fairly accurate scaling
relation applies after the spherical average despite the fact that the
departure from the spherical symmetry is quite visible in almost all
simulated halos.  A more realistic modeling of dark matter halos beyond
the spherical approximation is important in understanding various
observed properties of galaxy clusters and non-linear clustering of dark
matter.  In particular, the non-sphericity of dark halos is supposed to
play a central role in the X-ray morphologies of clusters, in the
cosmological parameter determination via the Sunyaev-Zel'dovich effect
and in the prediction of the cluster weak lensing and the gravitational
arc statistics (Bartelmann et al. 1998; Meneghetti et al. 2000, 2001).

Recently Jing \& Suto (2002) presented a detailed non-spherical modeling
of dark matter halos on the basis of a combined analysis of the
high-resolution halo simulations (12 halos with $N\sim 10^6$ particles
within their virial radius) and the large cosmological simulations (5
realizations with $N=512^3$ particles in a $100h^{-1}$Mpc boxsize).  The
density profiles of those simulated halos are well approximated by a
sequence of the concentric triaxial distribution with their axis
directions being fairly aligned:
%%%%%%%%%%%%%%%%%%%%%%%%%%%%%%%%%%%%%%%%%%%%%%%%%%%%%%%%%%%%%%%%%%%
\begin{eqnarray}
\label{eq:triaxial}
 R^2(\rho_{\rm s}) = \frac{X^2}{a^2(\rho_{\rm s})} 
+ \frac{Y^2}{b^2(\rho_{\rm s})} + \frac{Z^2}{c^2(\rho_{\rm s})} .
\end{eqnarray}
%%%%%%%%%%%%%%%%%%%%%%%%%%%%%%%%%%%%%%%%%%%%%%%%%%%%%%%%%%%%%%%%%%%
The origin of the coordinates is set at the center of mass of each
surface, and the principal vectors ${\bf a}$, ${\bf b}$ and ${\bf c}$
($a \le b \le c$) are computed by diagonalizing the inertial tensor of
particles in the isodensity surface $\rho=\rho_{\rm s}$.  Figure 6 plots
the density profiles measured in this way for individual halos as a
function of $R$, which indicates that equation (6) is still a good
approximation if the spherical radius $r$ is replaced by equation (7).
The application of the above triaxial modeling for the X-ray,
Sunyaev-Zel'dovich, and lensing data is in progress (Lee \& Suto 2003,
2004; Oguri, Lee \& Suto 2003).
%%%%%%%%%%%%%%%%%%%%%%%%%%%%%%%%%%%%%%%%%%%%%%%%%%%%%%%%%%%%%%%
\begin{figure}[htb]
\begin{center}
 \leavevmode\epsfxsize=10.0cm \epsfbox{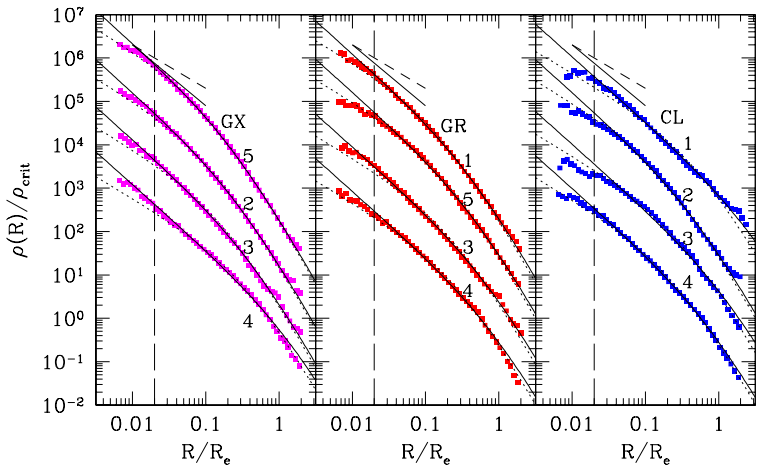} \caption{Radial
density profiles in our triaxial model of the simulated halos of galaxy
({\it left}), group ({\it middle}), and cluster ({\it right}) masses.
The solid and dotted curves represent fits to equation (6) with
$\alpha=1.5$ and 1.0, respectively.  For reference, we also show
$\rho(R) \propto R^{-1}$ and $R^{-1.5}$ in dashed and solid lines.  The
vertical dashed lines indicate the force softening length which
corresponds to our resolution limit.  For the illustrative purpose, the
values of the halo densities are multiplied by 1, $10^{-1}$, $10^{-2}$,
$10^{-3}$ from top to bottom in each panel (Jing \& Suto 2002).
\label{fig:dens_prof} }
\end{center} 
\end{figure}
%%%%%%%%%%%%%%%%%%%%%%%%%%%%%%%%%%%%%%%%%%%%%%%%%%%%%%%%%%%%%%%

\section{2000's: galaxies in cosmological hydrodynamic simulations} 

Although serious attempts to {\it create} galaxies phenomenologically
but directly from cosmological hydrodynamic simulations have been
initiated in early 1990's (e.g., Cen \& Ostriker 1992; Katz, Hernquist
\& Weinberg 1992), those resulting simulated galaxies are far from
realistic and there are still plenty of room for improvement. Thus this
is one of the most important, and yet quite realistic, goals for the
simulations in the new millennium, or hopefully in this decade.
%%%%%%%%%%%%%%%%%%%%%%%%%%%%%%%%%%%%%%%%%%%%%%%%%%%%%%%%%%%%%%%%%%%%%%%%%
\begin{figure}[thb]
\begin{center}
\leavevmode\epsfxsize=10cm \epsfbox{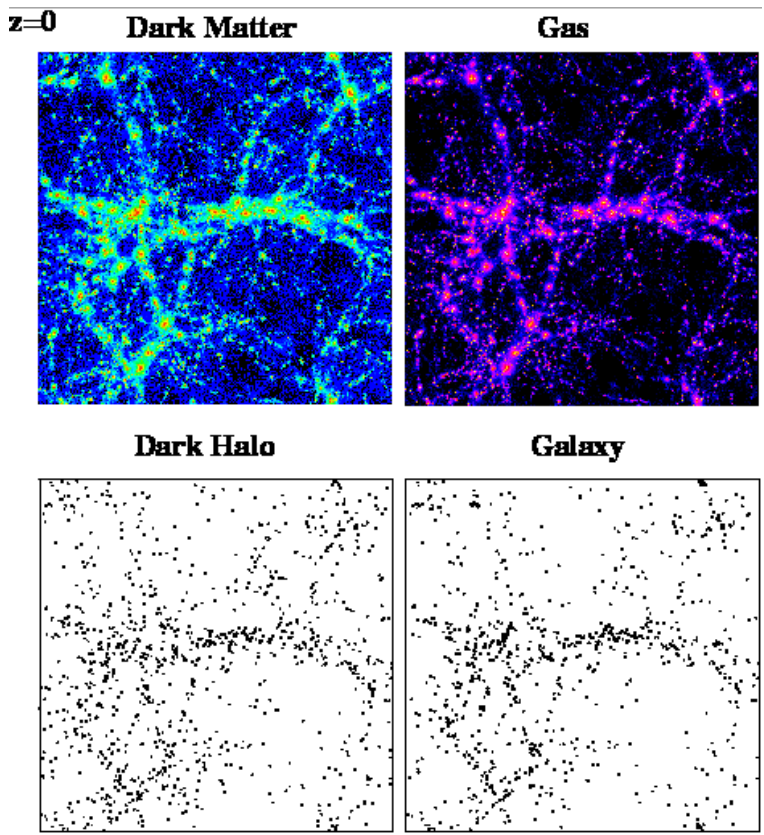} 
\caption{Distribution of gas particles, dark matter particles, galaxies
and dark halos in the volume of $75h^{-1}\times 75h^{-1}\times
30h^{-1}$Mpc$^3$ model at $z=0$. {\it Upper-right:} gas particles; {\it
Upper-left:} dark matter particles; {\it Lower-right:} galaxies; {\it
Lower-left:} dark halos (Yoshikawa et al. 2001) \label{fig:LCDM_Z00}}
\end{center}
\end{figure}
%%%%%%%%%%%%%%%%%%%%%%%%%%%%%%%%%%%%%%%%%%%%%%%%%%%%%%%%%%%%%%%%%%%%%%%%%

Let me show the result of Yoshikawa et al. (2001) for an example of such
approaches. They apply cosmological smoothed particle hydrodynamic
simulations in a spatially-flat $\Lambda$-dominated CDM model with
particular attention to the comparison of the biasing of dark halos and
simulated galaxies.  Figure 7 illustrates the distribution of dark
matter particles, gas particles, dark halos and galaxies at
$z=0$. Clearly galaxies are more strongly clustered than dark halos.  In
order to quantify the effect, we define the following biasing parameter:
%%%%%%%%%%%%%%%%%%%%%%%%%%%%%%%%%%%%%%%%%%%%%%%%%%%%%%%%%%%%%%%%%%%
\begin{equation}
 b_{\xi,i}(r)\equiv \sqrt{\frac{\xi_{ii}(r)}{\xi_{\rm mm}(r)}} ,
\end{equation}
%%%%%%%%%%%%%%%%%%%%%%%%%%%%%%%%%%%%%%%%%%%%%%%%%%%%%%%%%%%%%%%%%%%
where $\xi_{ii}(r)$ and $\xi_{\rm mm}(r)$ are two-point correlation
functions of objects $i$ and of dark matter, respectively.  Furthermore
for each galaxy identified at $z=0$, we define its formation redshift
$z_{\rm f}$ by the epoch when half of its {\it cooled gas} particles
satisfy our criteria of galaxy formation.  Roughly speaking, $z_{\rm f}$
corresponds to the median formation redshift of {\it stars} in the
present-day galaxies. We divide all simulated galaxies at $z=0$ into two
populations (the young population with $z_{\rm f}<1.7$ and the old
population with $z_{\rm f}>1.7$) so as to approximate the observed
number ratio of $3/1$ for late-type and early-type galaxies.
%%%%%%%%%%%%%%%%%%%%%%%%%%%%%%%%%%%%%%%%%%%%%%%%%%%%%%%%%%%%%%%%%%%%%%%%%
\begin{figure}[htb]
\begin{center}
\leavevmode\epsfxsize=7.5cm \epsfbox{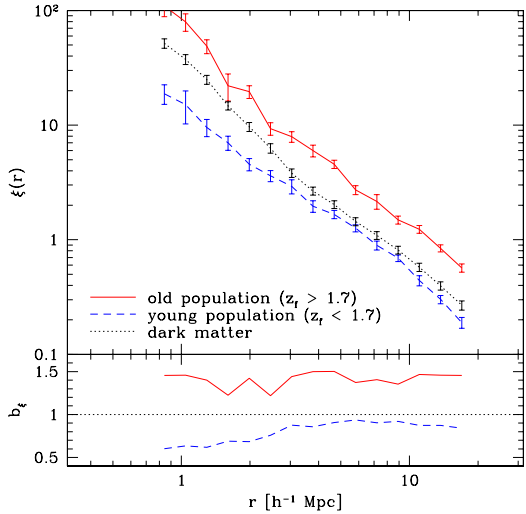} \caption{Two-point
correlation functions for the old and young populations of galaxies at
$z=0$ as well as that of dark matter distribution. The profiles of bias
parameters $b_{\xi}(r)$ for both of the two populations are also shown
in the lower panel (Yoshikawa et al. 2001). 
\label{fig:early_late_xi_bias}}
 \end{center}
\end{figure}
%%%%%%%%%%%%%%%%%%%%%%%%%%%%%%%%%%%%%%%%%%%%%%%%%%%%%%%%%%%%%%%%%%%%%%%%%

The difference of the clustering amplitude can be also quantified by
their two-point correlation functions at $z=0$ as plotted in Figure 8.
The old population indeed clusters more strongly than the mass, while
the young population is anti-biased. The relative bias between the two
populations $b^{\rm rel}_{\xi,{\rm g}} \equiv \sqrt{\xi_{\rm
old}/\xi_{\rm young}}$ ranges $1.5$ and 2 for
$1h^{-1}\mbox{Mpc}<r<20h^{-1}\mbox{Mpc}$, where $\xi_{\rm young}$ and
$\xi_{\rm old}$ are the two-point correlation functions of the young and
old populations. It is interesting to note that even this crude approach
is able to explain the morphological-dependence of bias, although still
in a rather quantitative manner, derived later by Kayo et al. (2004) for
SDSS galaxies. With the still on-going rapid progress of observational
exploration (e.g., Lahav \& Suto 2003 for a recent review on galaxy
redshift survey), understanding galaxy biasing as a function of galaxy
properties is definitely one of the unsolved important questions in
observational cosmology, and the present result indicates that the
formation epoch of galaxies plays a crucial role in the morphological
segregation.

\section{Distribution of dark baryons}

Finally let me briefly mention yet another possibility of tracing the
large-scale structure of the universe using the oxygen emission lines.
It is widely accepted that our universe is dominated by {\it dark}
components; 23 percent of dark matter, and 73 percent of dark energy
(e.g., Spergel et al. 2003). Furthermore, as Fukugita, Hogan \& Peebles
(1998) pointed out earlier, even most of the remaining 4 percent of the
cosmic baryons has evaded the direct detection so far, i.e., most of the
baryons is indeed {\it dark}.  Subsequent numerical simulations (e.g.,
Cen \& Ostriker 1999a, 1999b; Dav\'e et al. 2001) indeed suggest that
approximately 30 to 50 percent of total baryons at $z=0$ take the form
of the warm-hot intergalactic medium (WHIM) with $10^5 {\rm K}< T < 10^7
{\rm K}$ which does not exhibit strong observational signature.
%%%%%%%%%%%%%%%%%%%%%%%%%%%%%%%%%%%%%%%%%%%%%%%%%%%%%%%%%%%%%%%%%%%
\begin{figure}[h]
\begin{center}
\leavevmode\epsfysize=4.0cm \epsfbox{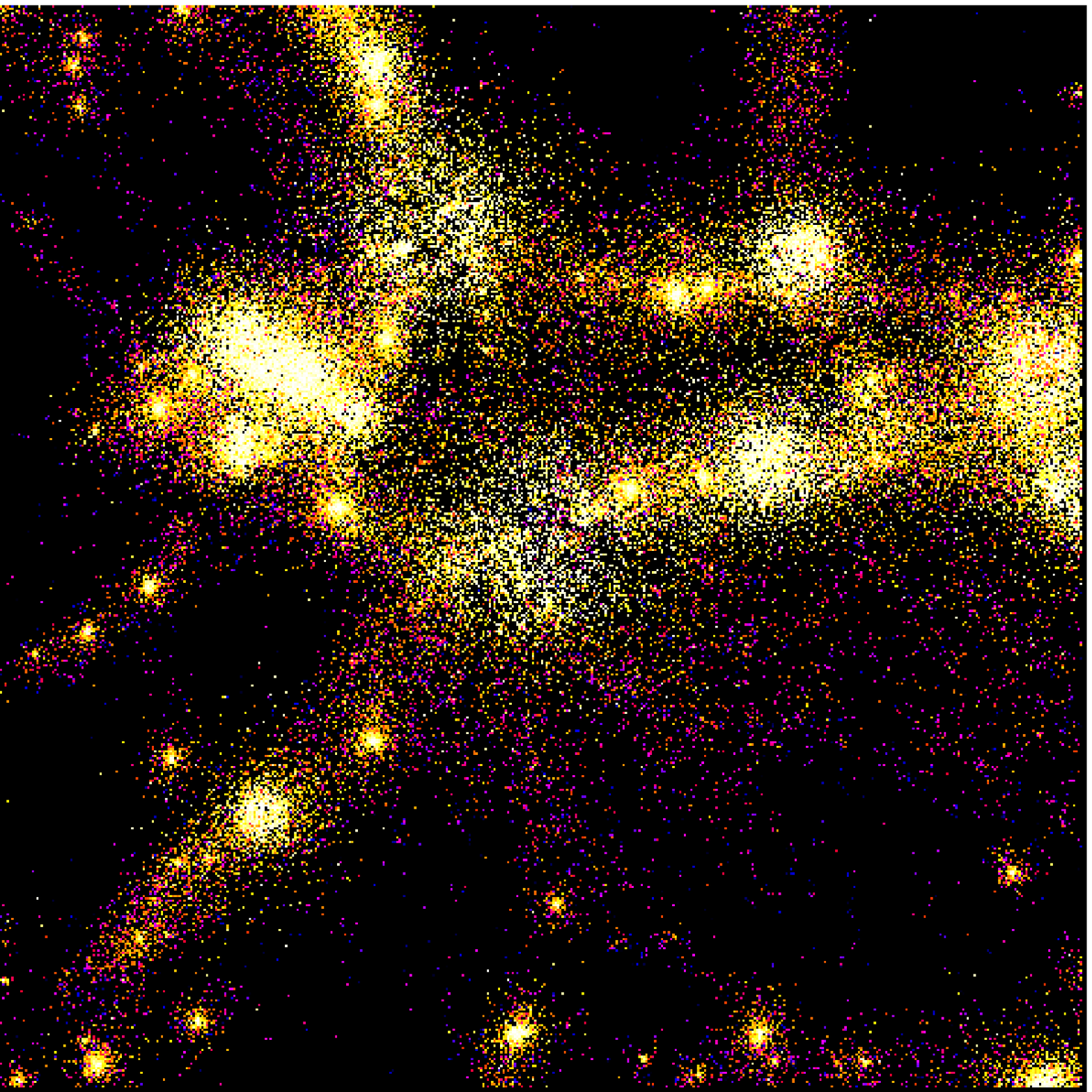}
\hspace*{0.2cm}
\leavevmode\epsfysize=4.0cm \epsfbox{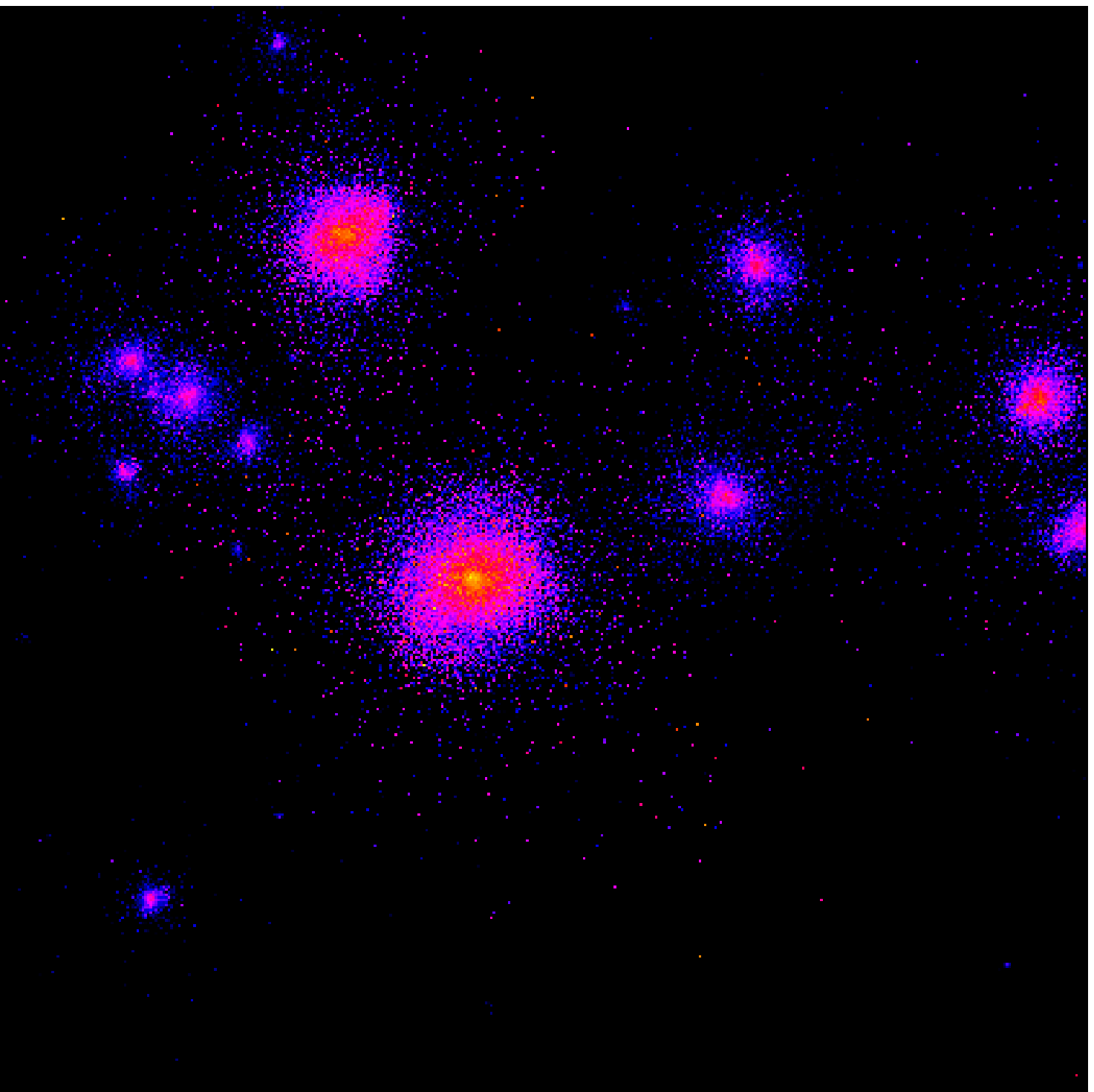}
\hspace*{0.2cm}
\leavevmode\epsfysize=4.0cm \epsfbox{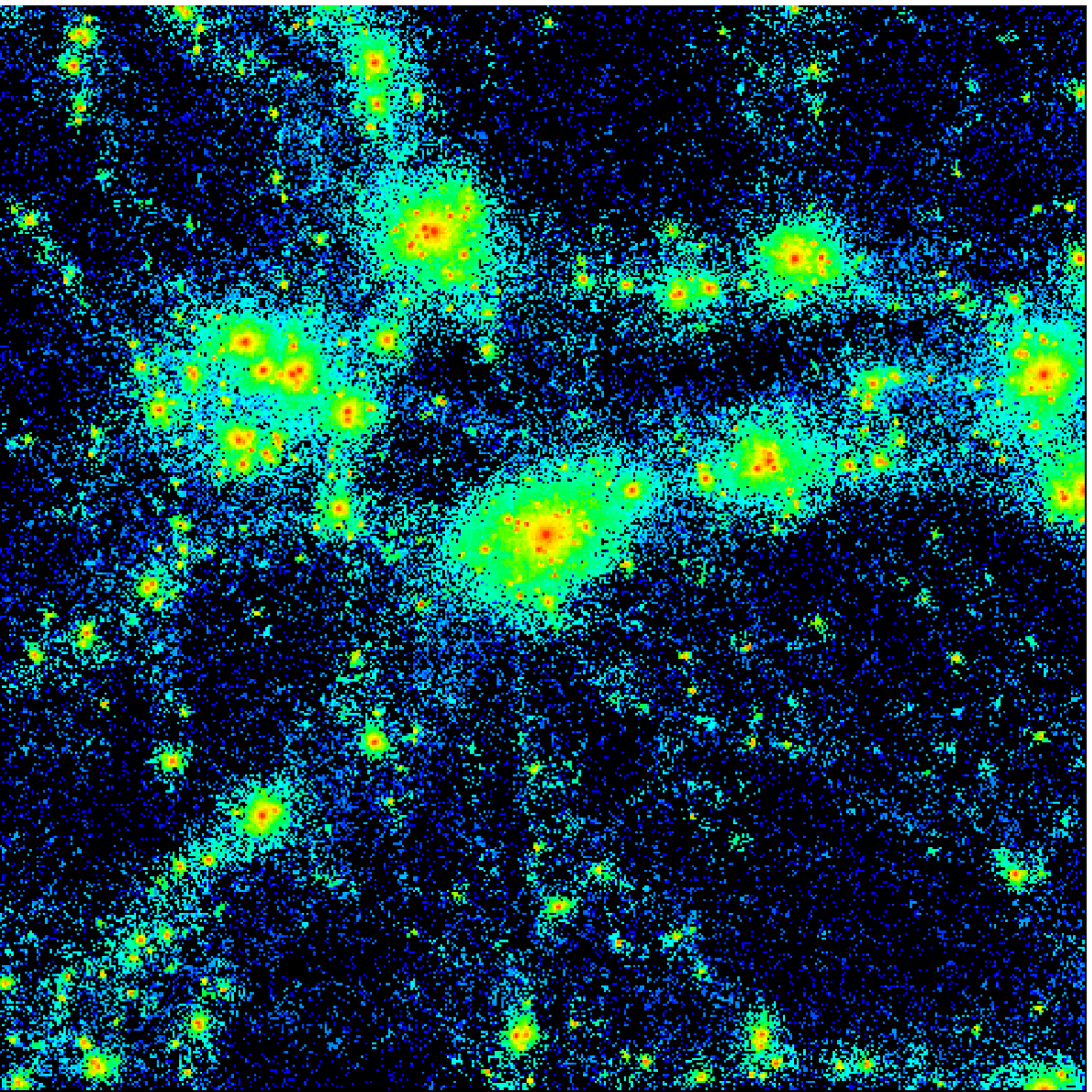}
\caption{Distribution of WHIM (left) compared with those of
hot intracluster gas (center) and dark matter (right).
The plotted box corresponds to $30h^{-1}$Mpc
 $\times 30h^{-1}$Mpc $\times 10h^{-1}$Mpc. 
\label{fig:whim}}
\end{center}
\end{figure}
%%%%%%%%%%%%%%%%%%%%%%%%%%%%%%%%%%%%%%%%%%%%%%%%%%%%%%%%%%%%%%%%%%%

Figure 9 compares the distribution of WHIM ($10^5 {\rm K}< T < 10^7 {\rm
K}$), hot intracluster gas ($T > 10^7 {\rm K}$), and dark matter
particles from cosmological smoothed-particle hydrodynamic simulations
(Yoshikawa et al. 2001). Clearly WHIM traces the large-scale filamentary
structure of mass distribution more faithfully than the hot gas which
preferentially resides in clusters that form around the knot-like
intersections of those filamentary regions.  In order to carry out a
direct and homogeneous survey of elusive dark baryons, we propose a
dedicated soft-X-ray mission, {\it DIOS} (Diffuse Intergalactic Oxygen
Surveyor; see Fig. 10).  The detectability of WHIM through O{\sc viii}
and O{\sc vii} emission lines via {\it DIOS} was examined in detail by
Yoshikawa et al. (2003) assuming a detector which has a large throughput
$S_{\rm eff}\Omega=10^2$ cm$^2$ deg$^2$ and a high energy resolution
$\Delta E = 2$ eV.  Their results are summarized in Figure 11; they
first create a light-cone output from the hydrodynamic simulation up to
$z=0.3$, compute the bolometric X-ray surface intensity map, select
several target fields and finally compute the corresponding spectra
relevant for the {\it DIOS} survey. The high-spectral resolution of {\it
DIOS} enables to identify the redshifts of several WHIMs at different
emission energies, i.e., Oxygen emission line tomography of the WHIMs at
different locations.
%%%%%%%%%%%%%%%%%%%%%%%%%%%%%%%%%%%%%%%%%%%%%%%%%%%%%%%%%%%%%%%%%%%
\begin{figure}[h]
\begin{center}
\leavevmode\epsfxsize=8.0cm \epsfbox{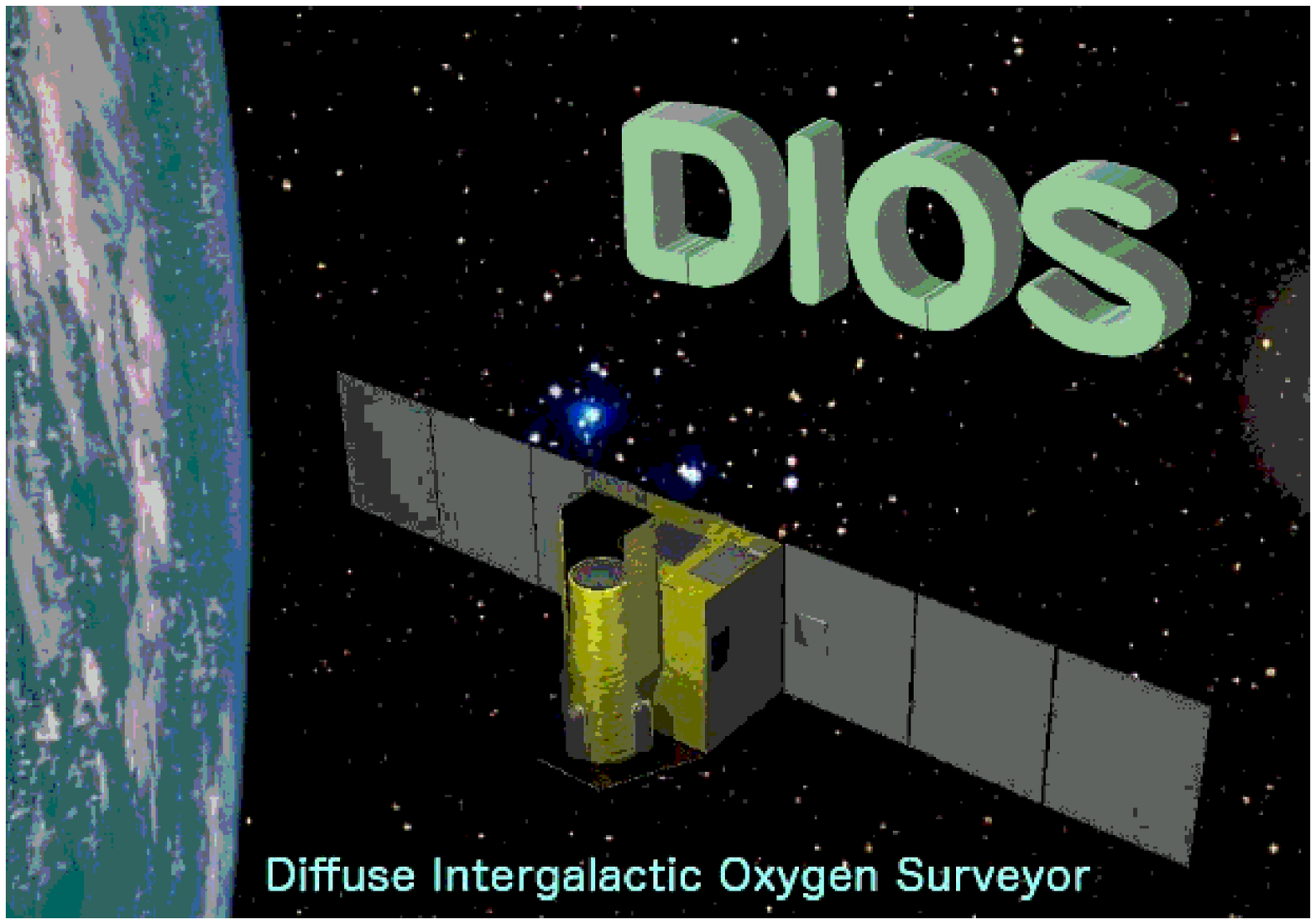} 
\caption{A dedicated soft X-ray mission to search for dark baryons 
via oxygen emission lines, {\it DIOS} (Diffuse Intergalactic Oxygen Surveyor).
\label{fig:dios_spacecraft}}
\end{center}
\vspace*{0.0cm}
%\end{figure}
%%%%%%%%%%%%%%%%%%%%%%%%%%%%%%%%%%%%%%%%%%%%%%%%%%%%%%%%%%%%%%%%%%%
%%%%%%%%%%%%%%%%%%%%%%%%%%%%%%%%%%%%%%%%%%%%%%%%%%%%%%%%%%%%%%%%%%%
%\begin{figure}[h]
\begin{center}
\leavevmode\epsfxsize=12.0cm \epsfbox{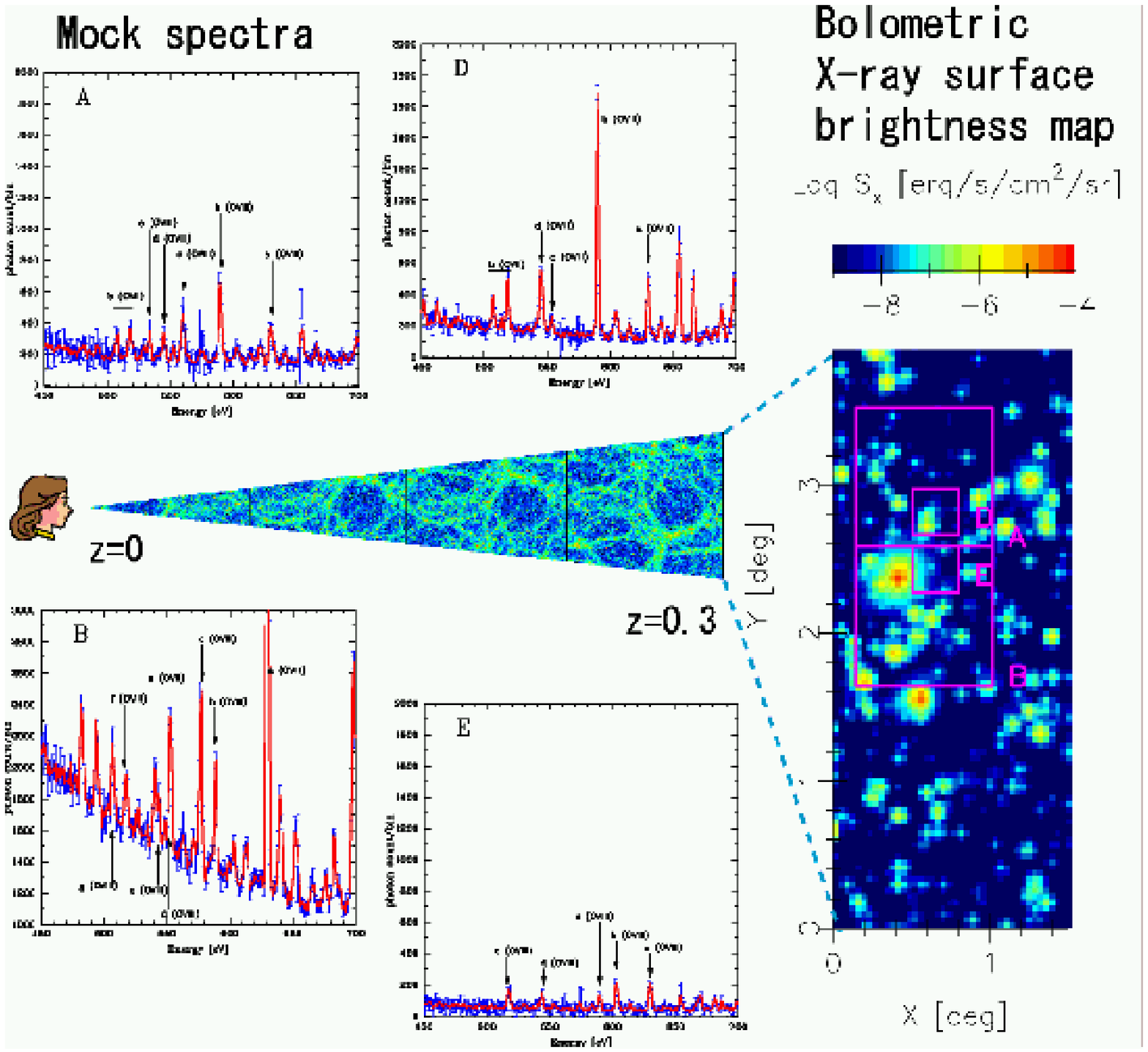} 
\caption{Mock spectra of WHIMs expected for {\it DIOS}.
\label{fig:oxygen}}
\end{center}
\end{figure}
%%%%%%%%%%%%%%%%%%%%%%%%%%%%%%%%%%%%%%%%%%%%%%%%%%%%%%%%%%%%%%%%%%%
\clearpage

They concluded that within the exposure time of $T_{\rm exp}=10^{5-6}$
sec {\it DIOS} will be able to reliably identify O{\sc viii} emission
lines (653eV) of WHIM with $T=10^{6-7}$ K and the overdensity of
$\delta=10^{0.5-2}$, and O{\sc vii} emission lines (561, 568, 574,
665eV) of WHIM with $T=10^{6.5-7}$ K and $\delta=10^{1-2}$.  The WHIM in
these temperature and density ranges cannot be detected with the current
X-ray observations except for the oxygen absorption features toward
bright QSOs. {\it DIOS} is especially sensitive to the WHIM with gas
temperature $T=10^{6-7}$K and overdensity $\delta=10-100$ up to a
redshift of $0.3$ without being significantly contaminated by the cosmic
X-ray background and the Galactic emissions. Fang et al. (2003) also
conducted a similar study and reached quite consistent conclusions. Thus
such a mission, hopefully launched in several years, promises to provide
a unique and important tool to trace the large-scale structure of the
universe via dark baryons.

\section{Conclusions}

It turned out that I was able to review the cosmological simulations
only for a time-scale of decades. Still the material may be heavily
biased, which I have to apologize for the organizers and possible
readers of these proceedings.  A millennium is definitely too long for
any scientist to make any reliable prediction for its eventual outcome.
Thus the number of simulation particles that I predicted in Introduction
might sound ridiculous, but in reality the progress in the new
millennium may be even more drastic that whatever one can imagine.  For
instance, it is unlikely that one still continues to use currently
popular particle- or mesh-based simulation techniques over next hundreds
of years. In that case, the number of particles may turn out to be a
totally {\it useless} measure of the progress or reliability of
simulations. Nevertheless I believe that a historical lesson that I
learned in preparing this talk will be still true even at the end of
this millennium; good science favors the prepared mind, not the largest
simulation at the time.

\bigskip
\acknowledgments

I thank Stephan Colombi, Gus Evrard, Takashi Hamana, Y.P.Jing, Atsushi
Taruya, Naoki Yoshida, and Kohji Yoshikawa for enjoyable
collaborations. Naoki Yoshida also encouraged me to plot Figure 1 in
order to illustrate the progress of cosmological N-body simulations.  I
am also grateful to Ed Turner for providing me a digitized version of
the first movie of his cosmological N-body simulations that I was able
to present in the talk. My presentation file for the symposium may be
found in the PDF format at
http://www-utap.phys.s.u-tokyo.ac.jp/{\textasciitilde}
suto/mypresentation\_2003e.  This research was supported in part by the
Grant-in-Aid for Scientific Research of JSPS.  Numerical computations
presented in this paper were carried out mainly at ADAC (the
Astronomical Data Analysis Center) of the National Astronomical
Observatory, Japan (project ID: yys08a, mky05a).  {\it DIOS} (Diffuse
Intergalactic Oxygen Surveyor) is a proposal by a group of scientists at
Tokyo Metropolitan University, Institute of Astronautical Sciences, the
University of Tokyo, and Nagoya University (P.I., Takaya Ohashi).

\clearpage
%%%%%%%%%%%%%%%%%%%%%%%%%%%%%%%%%%%%%%%%%%%%%%%%%%%%%%%%%%%%%%%%%%%%%

%%%%%%%%%%%%%%%%%%%%%%%%%%%%%%%%%%%%%%%%%%%%%%%%%%%%%%%%%%%%%%%%%%%%%
\end{document}